# ON MICROSCOPIC DESCRIPTION OF THE GAMMA – RAY STRENGTH FUNCTIONS

S. Kamerdzhiev, D. Voitenkov

Institute for Physics and Power Engineering, 249033 Obninsk, Russia

## Abstract

Using the Theory of Finite Fermi Systems, we obtained a non-magic nuclei generalization of the old theoretical results by J. Speth for magic nuclei dealing with the transitions between excited states and moment values of excited state. Such an extension is quite necessary for microscopic calculations of the gamma ray strength function. The comparison with the standard QRPA has shown that the modern many-body approach gives some new physics. The calculated value of the quadrupole moment of the excited $3_1^-$ state in $^{208}$Pb is agreed satisfactorily with the experiment. Some possible improvements of the theory are briefly discussed.

## 1. Introduction

As a rule, the gamma –ray strength function is used for the statistical studies of gamma transitions from the high-lying nuclear states near the nucleon separation energy.

One of the definitions of the gamma-ray strength function only includes the transitions between the ground and excited states [1]. Such a definition is true, in particular, for the so-called pygmy-dipole resonance which is being investigated at present very actively, both experimentally and theoretically, and is very important, as known, for the description of the radiative neutron capture; see, for example, a recent mini review [2]. Another definition of the gamma-ray strength function, and, in a sense, a conventional one, contains transitions between excited states [3].

In the context of quick and successful development of the microscopic nuclear structure methods and problems of gamma-ray strength function, discussed, for example, in [4, 5], it is necessary to consider modern microscopic feasibilities to calculate gamma-ray strength functions microscopically, which is especially important for the nuclei where there is no experimental information for phenomenological description, first of all, for neutron-rich nuclei. In particular, the question arises whether it is possible to compare microscopic and statistical approaches with the aim to understand limits of their applicability or the applicability of some approximations and hypotheses used. Thus, in order to try to understand the problem microscopically, it is necessary to calculate the transitions between excited states (the microscopic theory for the transitions between ground and excited states is very well developed; see, for example [6, 7, 8, 9]). This is the first motivation of our work. In addition, at present there is a huge number of experimental data in the field of conventional nuclear spectroscopy, which should be understood microscopically, see [10]. This is the second motivation of our work. In any case, it is of great interest to understand the novelty of the modern nuclear many-body theory in this field.

One of the modern nuclear structure microscopic approaches is the Theory of Finite Fermi Systems (TFFS) [6] and its extensions [7, 8, 9]. For the beginning, we would like to revive the old extended TFFS activity [7] at the modern level to calculate transitions between excited states, with each described within the Random Phase Approximation (RPA). In Section 2 we discuss the case of magic nuclei. In Section 3 we consider the case of non–magic nuclei and compare it with the usual Quasiparticle RPA. The first calculations for quadrupole moments of excited states are performed in Section 4.

## 2. Magic nuclei

We will use the microscopic extended TFFS in the old sense of this term [7, 8]. This version generalizes the standard TFFS [6] to calculate transitions between excited states and static moments of excited states for even-even magic nuclei. In [7], using the formalism of the six-point Green function for magic nuclei, the expression for the matrix element $M_{ss'}$ of the transition between excited states s→s′ was derived, with each described within RPA. It can be written in the graphic form [11, 12]

$$M_{ss'} = M_{ss'}^{(1)} + M_{ss'}^{(2)} \tag{1}$$

$$M_{ss'}^{(1)} = \sum_{123} V_{12}\, g_{31}^s\, g_{23}^{s'}\, A_{123}^{(1)} \tag{2}$$

$$M_{ss'}^{(2)} = \sum_{123} V_{12}\, g_{31}^{s'}\, g_{23}^{s}\, A_{123}^{(2)} \tag{3}$$

Here $g^s$ is the matrix element of the phonon creation amplitude with the phonon energy $\omega_s$, $V_{12}$ is the effective field which takes into account the effect of nuclear medium with respect to an external field $e_q V^0$ where $e_q$ is the local quasiparticle charge (for electric fields $e_q{}^p = 1$, $e_q = 0$) [6]. According to [7] the quantities $g^s$ and V should be calculated within the standard TFFS [6], that is, in fact, in RPA. The single–particle indexes for spherical nuclei are $\lambda \equiv \{n, j, l, m\} \equiv \{1\}$. The quantities $A^{(1)}$ and $A^{(2)}$ are given by

$$A_{123}^{(1)}(\omega_s, \omega_{s'}) = \int G_1(\varepsilon) G_2(\varepsilon + \omega) G_3(\varepsilon + \omega_s)\, d\varepsilon =$$

$$= \frac{(1-n_1)(1-n_2)n_3 - n_1 n_2 (1-n_3)}{(\varepsilon_{31} - \omega_s)(\varepsilon_{32} - \omega_{s'})} +$$

$$+ \frac{1}{\varepsilon_{12} + \omega} \left[ \frac{n_1(1-n_2)(1-n_3) - (1-n_1)n_2 n_3}{\varepsilon_{13} + \omega_s} - \frac{(1-n_1)(1-n_3)n_2 - (1-n_2)n_1 n_3}{\varepsilon_{23} + \omega_{s'}} \right] \tag{4}$$

$$A_{123}^{(2)}(\omega_s, \omega_{s'}) = A_{123}^{(1)}(-\omega_{s'}, -\omega_s)$$

where for spherical nuclei the single-particle indexes $\nu_1 = \{n,l,j\} \equiv 1$, $\varepsilon_{12} = \varepsilon_1 - \varepsilon_2$, $\omega = \omega_s - \omega_{s'}$, $n_1$ are occupation numbers ( $n_1 = 0,1$ for magic nuclei).

The case of the static moment of an excited state corresponds to $\omega = 0$. In particular, the quadrupole moment of the excited state s is as follows:

$$Q = \sqrt{\frac{16\pi}{5}} \langle I_s I_s | M | I_s I_s \rangle \tag{5}$$

with the quantity V in expressions (1), (2), (3)

$$V(\mathbf{r}) = V(r) Y_{20} \qquad (6)$$

and $I_s$ is the moment value of the excited state considered.

The main difference between expressions (1), (2), (3) and the RPA case is that instead of the external (or bare) field $e_q V^0$, which should be used in the RPA case, the effective field V is present. As it was stated in [7], the first term on the right side in (4) corresponds to the RPA case with $V = e_q V^0$. Note that this statement can be proved if one uses the property $n_i = n^2_i$, which is true for magic nuclei only. Thus, it is also of interest to find the QRPA case from the appropriate generalization of the approach under consideration (see Sect.3). Of course, it goes without saying that it is necessity to include pairing (non-magic nuclei) in the approach.

## 3. Non-magic nuclei

As known, for nuclei with pairing it is necessary to use four Green functions G, $G^h$, $F^{(1)}$, $F^{(2)}$ [6]

$$G_1(\varepsilon) = G_1^h(-\varepsilon) = \frac{u_1^2}{\varepsilon - E_1 + i\delta} + \frac{v_1^2}{\varepsilon + E_1 - i\delta}$$

$$F_1^1(\varepsilon) = F_1^2(\varepsilon) = -\frac{\Delta_1}{2E_1}\left[\frac{1}{\varepsilon - E_1 + i\delta} + \frac{1}{\varepsilon + E_1 - i\delta}\right] \qquad (7)$$

where $E_1 = \sqrt{(\varepsilon_1 - \mu)^2 + \Delta_1^2}$ and $\Delta_1$ is the gap value, which should be found from the solution of the gap equation, and to add the hp, pp and hh-channels in addition to the ph one of the RPA case, i.e. to consider the full set of QRPA equations and four effective fields V, $V^h$, $d^{(1)}$, $d^{(2)}$ [6]. Because the pp and hh channels usually give small contributions as compared with the ph and hp ones and create noticeable algebraic complications, here we omit the pp- and hh-channels and, correspondingly, the effective fields $d^{(1)}$ and $d^{(2)}$. In this approximation our matrix element $M_{ss'}$ contains 8 graphs:

$$M_{ss'} = \sum_i M_{ss'}^{(i)} = \qquad (8)$$

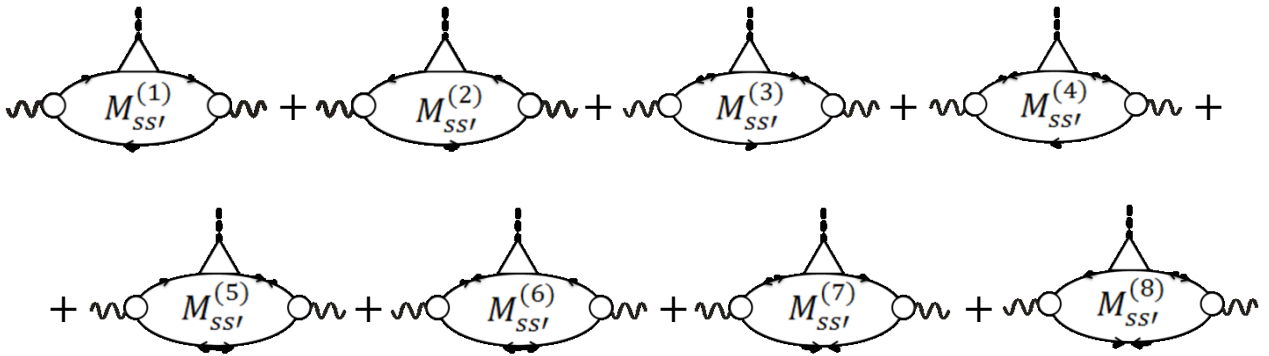

The analytical form for $M_{ss'}^{(1)}$ is given by

$$M_{ss'}^{(1)} = \sum_{123} V_{12} g_{31}^s g_{23}^{s'} A_{123}^{(1)} \qquad (9)$$

where

$$A_{123}^{(1)}(\omega_s, \omega_{s'}) = \int G_1(\varepsilon) G_2(\varepsilon + \omega) G_3(\varepsilon + \omega_{s'}) d\varepsilon =$$

$$= \frac{u_1^2 u_2^2 v_3^2}{(\omega_s + E_{13})(\omega_{s'} + E_{23})} - \frac{v_1^2 v_2^2 u_3^2}{(\omega_s - E_{13})(\omega_{s'} - E_{23})} + \qquad (10)$$

$$+ \frac{1}{\omega + E_{12}} \left( \frac{u_1^2 v_2^2 u_3^2}{E_{23} - \omega_{s'}} - \frac{u_1^2 v_2^2 v_3^2}{E_{13} + \omega_s} \right) + \frac{1}{\omega - E_{12}} \left( \frac{v_1^2 u_2^2 v_3^2}{E_{23} + \omega_{s'}} - \frac{v_1^2 u_2^2 u_3^2}{E_{13} - \omega_s} \right)$$

For the first graph $M^{(3)}$, which contains the Green function $F^{(1)}$ and $F^{(2)}$, we obtain the corresponding quantity $A^{(3)}{}_{123}$:

$$A_{123}^{(3)}(\omega_s, \omega_{s'}) = - \int G_1(\varepsilon) F_2(\varepsilon + \omega) F_3(\varepsilon + \omega_s) d\varepsilon =$$

$$= -\frac{\Delta_1 \Delta_2}{4 E_1 E_2} \left[ \frac{v_3^2}{(\omega_s + E_{13})(\omega_{s'} + E_{23})} + \frac{u_3^2}{(\omega_s - E_{13})(-\omega_{s'} + E_{23})} + \right. \qquad (11)$$

$$\left. + \frac{1}{\omega + E_{12}} \left( \frac{v_3^2}{\omega_s + E_{13}} - \frac{u_3^2}{-\omega_{s'} + E_{23}} \right) + \frac{1}{\omega - E_{12}} \left( \frac{v_3^2}{-\omega_{s'} - E_{23}} - \frac{u_3^2}{\omega_s - E_{13}} \right) \right]$$

Here $E_{12} = E_1 + E_2$, $u_1^2 = \frac{E_1 + (\varepsilon_1 - \mu)}{2 E_1} = 1 - v_1^2$

The other quantities $A^{(i)}$ are obtained from the relations:

$$A_{123}^{(2)}(\omega_s, \omega_{s'}) = A_{123}^{(1)}(-\omega_{s'}, -\omega_s), \qquad A_{123}^{(4)}(\omega_s, \omega_{s'}) = A_{123}^{(3)}(-\omega_{s'}, -\omega_s),$$
$$\qquad (12)$$
$$A_{123}^{(6)}(\omega_s, \omega_{s'}) = A_{123}^{(5)}(-\omega_{s'}, -\omega_s), \qquad A_{123}^{(8)}(\omega_s, \omega_{s'}) = A_{123}^{(7)}(-\omega_{s'}, -\omega_s)$$

For the case $\omega = 0$ the quantity $M_{ss'}$, like in the RPA case, contains the sum $\Sigma A^{(i)}$ of eight terms.

At present there is the QRPA case for $M_{ss'}$ which was studied in work [13]. To compare our results with this case, in each of the $A^{(i)}$ we only take two terms which contain denominators $(E_{13} - \omega_s)(E_{23} - \omega_{s'})$ and $(E_{13} + \omega_s)(E_{23} + \omega_{s'})$ and call it $\Sigma[A(i)]_0$. Using relatively long algebra one can obtain:

$$\sum_i \left[ A_{123}^{(i)} \right]_0 = \left[ \frac{1}{(\omega_s + E_{13})(\omega_s + E_{23})} + \frac{1}{(\omega_s - E_{13})(\omega_s - E_{23})} \right] \times \qquad (13)$$

$$\times \left[ u_1^2 u_2^2 v_3^2 - v_1^2 v_2^2 u_3^2 + \frac{\Delta_1 \Delta_2}{4 E_1 E_2}(u_3^2 - v_3^2) + \frac{\Delta_1 \Delta_3}{4 E_1 E_3}(u_2^2 - v_2^2) + \frac{\Delta_2 \Delta_3}{4 E_2 E_3}(u_1^2 - v_1^2) \right]$$

The second square bracket in (13) coincides with the factor $v_{12}^- u_{23}^+ u_{31}^+$ in Refs. [13, 14] so that our quantity (13) is proportional to the factor $v_{12}^-(\psi_{23}\psi_{31} + \varphi_{23}\varphi_{31})$ in [13], which determines the appropriate matrix element within the QRPA ($\psi$ and $\varphi$ are phonon amplitudes). Thus, the terms containing factors $(\omega \pm E_{12})^{-1}$ in (10), (11), (12) are added in our approach and, therefore, generalize the standard QRPA. The second generalization of the QRPA is the appearance of the $\omega$ dependent effective field $V(\omega)$, with $\omega = \omega_s - \omega_{s'}$ instead of the external field $e_q V^0$, which does not depend on $\omega$.

## 4. Some calculations for the magic nuclei

Here we consider the case of $\omega = 0$ for magic nuclei and calculate quadrupole moments of excited states in Pb[208]. Then the quantities $A^{(1)}$ and $A^{(2)}$ in (1), (2), (3), (4) depend on $\omega_s$ only and the matrix element $M_{ss}$ contains the sum $\left[A^{(1)}_{123}(\omega_s) + A^{(1)}_{123}(-\omega_s)\right]$.

It is easy to see from (4) that for the case of $\omega = 0$ there is an uncertainty at 1=2 in eq. (4) of the kind 0/0. For this reason we should consider the case $A_{113}$ separately:

$$A^{(1)}_{113} = \frac{n_3 - n_1}{(\varepsilon_{31} - \omega_s)^2}, \quad A^{(2)}_{113} = \frac{n_3 - n_1}{(\varepsilon_{31} + \omega_s)^2} \tag{14}$$

We will consider the quantities V and $g^s$ in the following form :

$$V(\vec{r}) = V(r) Y_{LM}(\theta,\varphi), \quad g^s(\vec{r}) = g(r) Y_{l_s M_s}(\theta,\varphi) \tag{15}$$

that is for simplicity we neglect the spin components of these quantities, which is reasonable for the electric fields considered.

After summation over quantum numbers $m_1$, $m_2$, $m_3$ in (1), (2), one can obtain the following for the matrix element $M_{ss}$ :

$$M_{ss} = (-1)^{I_s+L} \begin{pmatrix} I_s & L & I_s \\ -M_s & M & M_s \end{pmatrix} \sum_{123} \begin{Bmatrix} I_s & L & I_s \\ j_2 & j_3 & j_1 \end{Bmatrix} \times$$

$$\times \langle 1\|V\|2\rangle\langle 3\|g^s\|1\rangle\langle 2\|g^s\|3\rangle \left[A^{(1)}_{123}(\omega_s) + A^{(1)}_{123}(-\omega_s)\right] \tag{16}$$

Here index 1 stands for $\nu_1 \equiv \{1\} = \{n_1, l_1, j_1\}$, the reduced matrix elements and the quantity $(A^{(1)} + A^{(2)})$ are given by

$$\langle 1\|V\|2\rangle = [V(r)]_{12}\langle 1\|Y_L\|2\rangle$$

$$\langle 3\|g^s\|1\rangle = [g^s(r)]_{31}\langle 3\|Y_{l_s}\|1\rangle$$

$$\langle 1\|Y_L\|2\rangle = \frac{(-1)^{j_2+L+1/2}}{2\sqrt{\pi}} [(2l_1+1)(2L+1)(2l_2+1)(2j_1+1)(2j_2+1)]^{1/2} \times \tag{17}$$

$$\times \begin{pmatrix} l_1 & L & l_2 \\ 0 & 0 & 0 \end{pmatrix} \begin{Bmatrix} I_s & L & I_s \\ j_2 & j_3 & j_1 \end{Bmatrix}$$

$$A^{(1)}_{123}(\omega_s) + A^{(1)}_{123}(-\omega_s) = (1-\delta_{12})\frac{2}{\varepsilon_{12}}\left[\frac{n_{13}\varepsilon_{13}}{\varepsilon_{13}^2 - \omega_s^2} - \frac{n_{23}\varepsilon_{23}}{\varepsilon_{23}^2 - \omega_s^2}\right] + \delta_{12}\frac{2n_{31}(\varepsilon_{31}^2 + \omega_s^2)}{(\varepsilon_{31}^2 - \omega_s^2)^2} \tag{18}$$

where $[V(r)]_{12}$ and $[g^s(r)]_{31}$ are radial integrals.

For the beginning, in our calculations we used the following approximations [15]:

$$V(r) = e_{eff}V^0, \quad e^p_{eff} = 2, \quad e^n_{eff} = 1 \qquad (19)$$

and the Bohr-Mottelson model for $g^s(r)$:

$$g^s(r) = \frac{\beta_s}{\sqrt{2I_s + 1}} r \frac{\partial U}{\partial r} \qquad (20)$$

where U is the single-particle potential.

To calculate the quadrupole moment value we used formulas (5) and (6). Summation over 1,2,3 in (15) was performed within two shells above and two shells below the Fermi energy both for protons and neutrons.

For $^{208}$Pb we obtained the following value of the quadrupole moment of the excited state $3^-_1$ with the energy $\omega_s = 2.6\ MeV$ and $\beta_s = 0.12$
$$Q_{th}(3^-_1) = -0.26\ e \cdot b$$

The experiment value $Q_{exp}(3^-_1) = -0.34 \mp 15\ e \cdot b$ [10].

We do not think that the consistent use of the RPA approach to calculate $V$ and $g^s$ instead of our approximations (18), (19), (20) can improve considerably the numerical result obtained. There are other physical reasons to improve it. The first is the use of the coordinate representation within the Green function method which allow us to consider the single-particle continuum exactly and omit the questions about the size of the summation limits in our formulas. The second reason is an account for the effects of the so-called tadpole. These effects were introduced and calculated for other problems in the works by the Kurchatov institute group long ago, see [16]. However, they should give a contribution to the problems which have been considered here.

## Conclusion

In this work, we have tried to outline modern feasibilities of the many-body nuclear theory to calculate transitions between excited states and, as a specific case, moment values of the excited state. It has been assumed that these excited states are described within the (Q)RPA. In fact, this assumption has been done for the simplicity and based on theoretical results available; it is possible to consider a more general approach, see for example [14,16].

For magic nuclei, more simple formulas (18) have been derived here than in the pioneering work by Speth [7] and the case of the appearance of the uncertainty has been considered. For non-magic nuclei, an appropriate generalization of the results [7] has been performed and compared with the QRPA approach [13]. This generalization allowed us to directly demonstrate a new physics due to the use of the modern many-body theory.

Our first and approximate calculation of the quadrupole moment of the $3^-_1$-level in $^{208}$Pb gave a reasonable agreement with the experiment [10] which has relatively large errors. As the approximations used are rather reasonable, we do not think that consistent use of the RPA will give a noticeable change of our numerical result. Of course, these approximations are not true for unstable nuclei which are studied very actively at present. For these nuclei, the self-consistent approaches are necessary and, for example, the extended theory of finite Fermi systems [17] gives some appropriate possibilities.

We think that the two above-mentioned physical effects (coordinate representation and tadpole effects) are very interesting and should be important numerically for calculation of transitions between excited states and moments of excited states, especially in neutron-rich nuclei. These effects should be considered first of all for these problems.

We are grateful to J. Speth ant B.A. Tulupov for enlightening and useful discussions.

The work was partly supported by the grants DFG 436RUS113/994/0-1 and RFBR 09-02-91352.